\documentclass{moriond}

\def\be{\begin{equation}}
\def\ee{\end{equation}}
\def\bea{\begin{eqnarray}}
\def\eea{\end{eqnarray}}

\newcommand{\eq}[1]{\eqref{eq:#1}}
\newcommand{\fig}[1]{Fig.~\ref{fig:#1}}

\def\s0#1#2{\mbox{\small{$ \frac{#1}{#2} $}}}
\def\0#1#2{\frac{#1}{#2}}
\usepackage{amstext,amssymb,amsmath}

\newcommand{\Photo}{}

\begin{document}
\vspace*{4cm}

\title{Vacuum Stability as a Guide for Model Building}

\author{Gudrun~Hiller,${}^{a,b}$ Tim~H\"ohne,${}^a$ \underline{Daniel~F.~Litim},${}^b$ and Tom~Steudtner\,${}^{a,c}$}
\address{${}^a$Department of Physics, TU Dortmund, Otto-Hahn-Str.4, D-44221 Dortmund, Germany\\
${}^b$Department of Physics and Astronomy, University of Sussex, Brighton, BN1 9QH, U.K.\\
${}^c$Department of Physics, University of Cincinnati, Cincinnati, OH 45221, USA}

\maketitle

\abstracts{
We explain why  vector-like fermions are natural candidates to lift  the Standard Model vacuum instability.
Results are further discussed  from the viewpoint of   criticality.
Several models allow for  vector-like quarks and leptons  in the TeV-range which can be searched for at the LHC.}

\section{Introduction}

The discovery of the Higgs  particle  together with theoretical precision calculations  evidenced the instability of the   standard model (SM) vacuum.\,\cite{CMS:2012qbp,ATLAS:2012yve,Degrassi:2012ry}  
While  a theory of nature with a decaying ground state  would be  unacceptable, vacuum metastability due to a lifetime sufficiently large compared to the age of the universe has become a widely accepted narrative. Further, the continuing success of the SM in the LHC era,  
with only a few anomalies and the absence of  clear new physics signatures  at colliders 
or elsewhere, calls for new   ideas and directions in model building. 
In this contribution, we promote the quest for vacuum stability 
 into a primary model building task.\cite{Hiller:2022rla}  
The rationale for this is that while the onset of the SM instability is a high energy effect, unattainable by present or planned future colliders,  its  existence   alone does not point towards a specific  scale for new physics. Therefore,  solutions could emerge from novel phenomena at any scale below the Planck scale,  and potentially  as low as a few TeV. 

.\section{Vacuum Stability}

\begin{figure*}
    \centering
    \includegraphics[trim={2cm 0 2cm 0},clip,  width=.9\columnwidth]{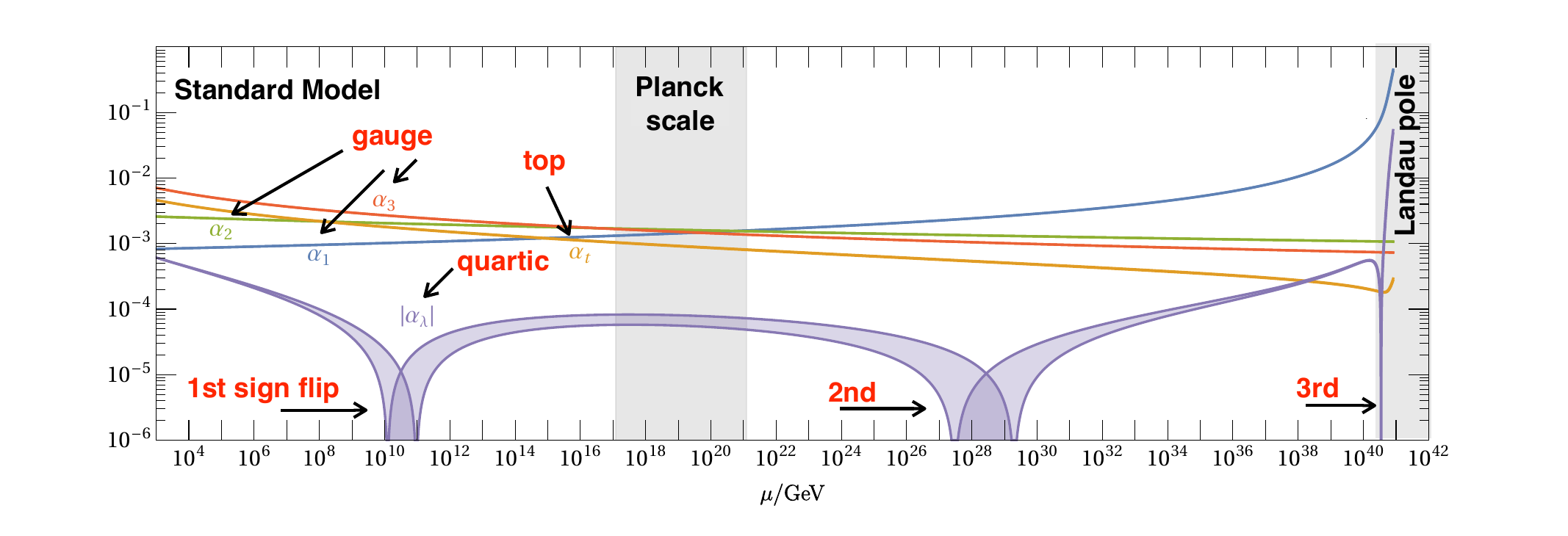}
   \vskip-.5cm
     \caption{Shown is the Standard Model 3-loop running of the Higgs quartic, top Yukawa, and gauge couplings   above  TeV energies.     The  vacuum becomes unstable ($\mu \approx 10^{10}$ GeV) prior to the Planck scale  (center gray band). Subsequently, and ignoring quantum gravity effects,  the steady growth of the hypercharge coupling  re-instates stability ($\mu \approx 10^{29}$ GeV) much before perturbativity, stability, and predictivity are ultimately lost  at a Landau pole  ($\mu \approx 10^{41}$ GeV). Bands indicate   a $1\sigma$ uncertainty in the  top pole mass.} 
    \label{fig:pSM}
\end{figure*}

We begin by taking stock of vacuum stability in the SM. 
To that end, we study the 3-loop  running of SM couplings up to the Planck scale and beyond.\,\cite{Hiller:2022rla,Litim:2020jvl,Hiller:2019mou}
We  introduce   the $U(1)_Y \times SU(2)_L \times SU(3)_c$ gauge couplings $g_{\ell}\ (\ell=1,2,3)$, the top and bottom Yukawa interactions $y_{t,b}$, and the Higgs quartic $\lambda$, all normalized in units of  loop factors, and write them as
 \begin{equation} \label{eq:alpha}
  \alpha_\ell = \frac{g_\ell^2}{(4\pi)^2}, \quad \alpha_{t,b} = \frac{y_{t,b}^2}{(4\pi)^2}, \quad \alpha_\lambda = \frac{\lambda}{(4\pi)^2}\,.
\end{equation}
 SM initial conditions (central values) are determined at the reference scale $\mu_0 =$~1~TeV. 
The uncertainties  in the initial values  due to the strong gauge coupling, Higgs and  $W$ mass are quantitatively irrelevant. 
The dominant source of uncertainty  originates from the determination of the top mass
$m_t = 172.76  \pm 0.30~{\rm GeV},$\,\cite{ParticleDataGroup:2020ssz}   which is indicated in~\fig{pSM} by a $1\sigma$~uncertainty band for all couplings.
Due to its smallness, the bottom Yukawa $\alpha_b(\mu)$ is  not displayed even though it is retained in the numerics.
 
  Unsurprisingly, \fig{pSM} confirms  that SM couplings run slowly. 
Most notably, however, and within uncertainties, the Higgs quartic invariably displays a  sign flip around $\mu\approx 10^{10}$~GeV, signaled by  a downward spike, and indicating the onset of vacuum instability. 
Stability up to  the Planck scale would require that the top mass deviates by more than $3\sigma$ from its presently determined central value. Hence, a negative value for the Higgs quartic at the Planck scale 
\begin{equation}\label{negquart}
\alpha_\lambda\big|_{\mu=M_{\rm Pl}}\approx - 10^{-4}
\end{equation}
and the possibility of an unstable ``great desert'' should be taken for real.
Curiously, 
extending the flow beyond the Planck scale, we observe that the Higgs becomes stable   again ($\mu_{\rm stab}\approx 10^{10} M_{\rm Pl}$), largely triggered by the mild but continued growth of the hypercharge coupling.
At much higher energies ($\mu_{\rm Landau}\approx 10^{13} \mu_{\rm stab}$), however, stability, perturbativity, and predictivity are ultimately lost, in this order and for good, and the SM as we know it comes to an end. 
Hence, new effects  are required to stabilise the vacuum, either  straight out of quantum gravity, 
or  from  particle physics  via new matter fields or interactions.

\section{Model Building Directions}

Vacuum stability  can be achieved through  BSM effects, as long as these  enhance the Higgs quartic sufficiently strongly.\,\cite{Hiller:2022rla} 
Minimally, this can be done  by introducing new particles which only couple to the SM gauge fields (``gauge portals''). 
 One may also  introduce new interactions involving the Higgs and the BSM fields such as new Yukawas (``Yukawa portal'') or new quartics (``Higgs portal''), or  other; see~\cite{Hiller:2019mou,Bissmann:2020lge,Bause:2021prv} for recent examples. Gauge portals 
 only modify parameters of the SM beta functions and lead to mild effects. Yukawa, Higgs, and other portals
 add new interactions, and   thereby modify the running of couplings more significantly.\,\cite{Hiller:2022rla,Hiller:2019mou,Bissmann:2020lge,Bause:2021prv}
  
The main idea for the gauge portal mechanism \cite{Hiller:2022rla}  could not be any simpler:~add $N_F$ new vector-like fermions   (VLF) to the SM, with charges $(Y_F,d_2,d_3)$  under   the $U(1)_Y \times SU(2)_L \times SU(3)_c$  gauge group.   By design, any of these SM extensions  are free of gauge anomalies, and allow for  Dirac mass terms $M_F$,
\begin{equation}
\mathcal{L}_{\rm BSM} \supset \bar \psi\left(  i  \!  \! \not \! \! D -M_F\right)\psi \,.
\end{equation}
From the viewpoint of the renormalisation group (RG), the primary  effect  is that the new particles
modify the running of gauge couplings. Specifically,  gauge beta functions   $\beta_i \approx - B_i\,\alpha_i^2$ have  modified one-loop  coefficients  
    $ B_1 = -\s0{41}3 - \delta B_1\,, 
     B_2 =  \s0{19}3 - \delta B_2\,,$ and 
     $B_3 =  14 - \delta B_3\,,$
 with positive VLF contributions 
       $\delta B_1 = \s083 N_F \, d_2\, d_3 \,Y_F^2$ and      $\delta B_{2,3} =\s083 N_F \, d_{3,2} \, S_2(d_{2,3})$ 
 in terms of their hypercharge $Y_F$ and Dynkin indices $S_2(d_{2,3})$ under  $SU(2)_L$ and $SU(3)_c$.

      Let us briefly explain how   new matter fields modify the running of the Higgs quartic. For simplicity, we take the BSM fermion mass as  the matching scale $\mu_0=M_F$ to SM running. VLFs then decouple at  scales below their own mass, and contribute as if they were massless at scales above. Subleading threshold corrections are neglected. For RG scales $\Lambda > \mu_0$, $\delta B_{i} \geq 0$ implies that gauge couplings  take values larger or equal to their  SM values,
      \begin{eqnarray}
         \alpha_\ell(\Lambda) - \alpha^\text{SM}_\ell(\Lambda)&\ge& 0\,. \label{gauge}
\end{eqnarray}
For the top Yukawa,  
we observe from   $\beta_t \approx \alpha_t \left[ 9\,\alpha_t - \tfrac{17}6\, \alpha_1 - \tfrac92\, \alpha_2  - 16 \, \alpha_3\right]$  that all gauge couplings contribute negatively  to its leading order running. Together with \eqref{gauge}, we conclude that the top Yukawa becomes smaller than in the SM,
      \begin{eqnarray}       \alpha_t(\Lambda) - \alpha^\text{SM}_t(\Lambda)&<&0\,.\label{top}
 \end{eqnarray}
Finally, we turn  to the Higgs quartic coupling $\alpha_\lambda$. Given that its value  is much smaller than the   top Yukawa and gauge  couplings, \fig{pSM}, its running is primarily driven by the inhomogeneous terms, 
    $\beta_\lambda \approx \tfrac{3}{8}\left[\alpha_1^2 + 2\, \alpha_1 \alpha_2 + 3\, \alpha_2^2\right] - 6\, \alpha_t^2 \,.$
Most notably, the gauge and top Yukawa couplings contribute with opposite signs, which in view of \eqref{gauge} and \eqref{top} means that they all pull into the same direction.  Overall, the Higgs quartic is invariably enhanced over its SM value,
\begin{eqnarray}  \alpha_\lambda(\Lambda) - \alpha^\text{SM}_\lambda(\Lambda)&>&0\,.\label{quartic}
\end{eqnarray}
This is the gauge portal mechanism. We conclude that vector-like fermions are natural candidates to  stabilise the electroweak vacuum.
It then remains to be seen  whether the  uplift \eqref{quartic} is sufficient to offset metastability \eqref{negquart}. 
To leading logarithmic accuracy, we find
\begin{eqnarray}\label{eq:quartic-enhance}
        \alpha_\lambda(\Lambda) - \alpha^{\rm SM}_\lambda(\Lambda) &\approx&
         \ \ \s038 \alpha_1^2(\mu_0)\left[\alpha_1(\mu_0) + \alpha_2(\mu_0)\right] \delta B_1 \, \ln^2\left({\Lambda}/{\mu_0}\right) \nonumber \\
        && + \s038 \alpha_2^2(\mu_0)\left[\alpha_1(\mu_0) + 3\alpha_2(\mu_0)\right] \delta B_2 \, \ln^2\left({\Lambda}/{\mu_0}\right) 
        \\&& + 32\,\alpha_t^2(\mu_0)\,\alpha_3^2(\mu_0)\,\delta B_3 \,\ln^3\left({\Lambda}/{\mu_0}\right) +{\rm subleading}\,.\nonumber 
\end{eqnarray}
A few comments are in order.
$(i)$  The leading impact from the hypercharge and weak isospin interactions, characterized by the terms   $\propto \delta B_{1,2}  \ln^2\left({\Lambda}/{\mu_0}\right)$, originates from the direct uplift of the Higgs quartic  at 2-loop level and leading logarithmic accuracy.
$(ii)$   Since the Higgs is colourless,
 the leading impact from  strong  interactions  is channeled through the top Yukawa coupling, and $\propto \delta B_3  \ln^3\left({\Lambda}/{\mu_0}\right)$ instead.
The additional loop suppression
may very well be compensated by the sizeable    prefactor, also depending on VLF masses, gauge charges and multiplicities. 
$(iii)$   Since the leading loop coefficients of the scalar and top Yukawa beta functions have not changed, the modified running of gauge couplings implies that $\alpha_\lambda$  approximately runs along the SM trajectory  $\alpha_\lambda(\mu)\approx \alpha^{\rm SM}_\lambda(\mu_{\rm SM})$, though with an altered ``RG velocity''. 
From \eqref{gauge}  we have $\mu_{\rm SM}(\mu) \gtrsim \mu$ for the weak and strong  portals,  
effectively uplifting  $\alpha_\lambda$.
For the hypercharge portal we find that $\mu_{\rm SM}(\mu)< \mu$ instead, and  the  $\alpha_\lambda$ trajectory comes out as a ``squeezed'' version of the SM one. The general case is a combination of these two effects.
 $(iv)$ In any gauge portal extension, the Higgs quartic is naturally bounded from below by its most negative value achieved along the SM trajectory,  and, incidentally, given by its value at the Planck scale, \eqref{negquart}.
 
\section{Gauge Portals at Work}

\begin{figure*}
    \centering
    \includegraphics[trim={2cm 0 2cm 0},clip,
    width=.9\columnwidth]{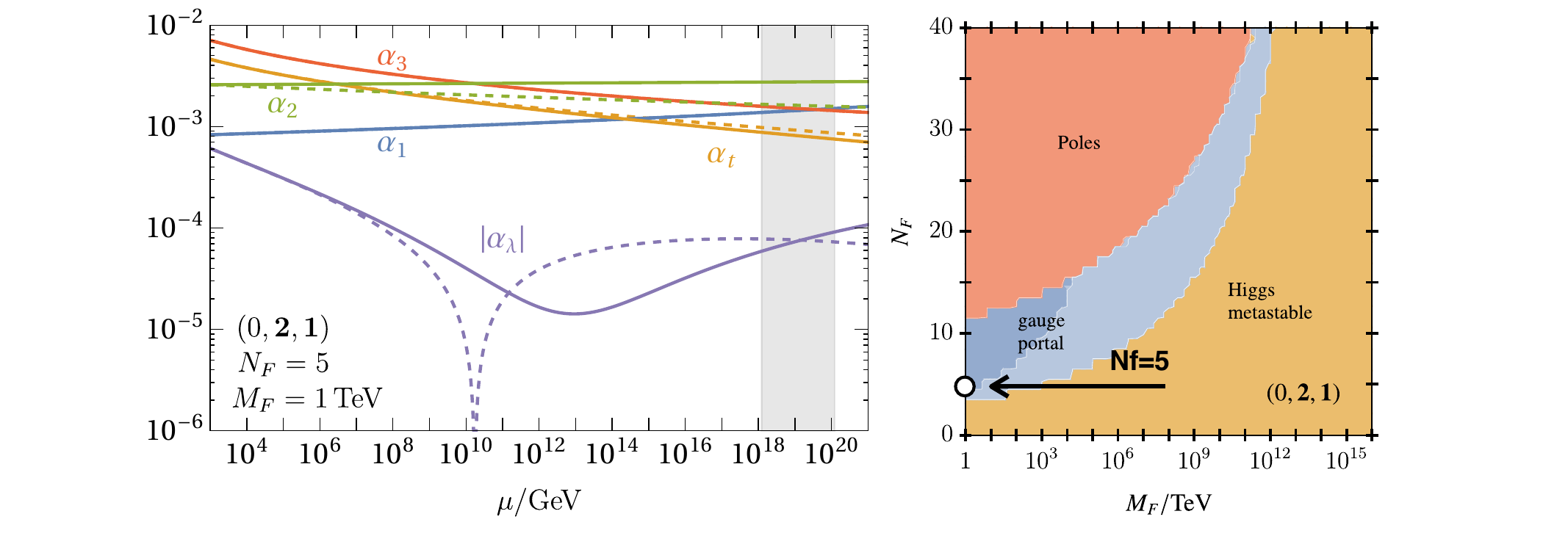}
\vskip-.5cm
    \caption{ Illustration of the weak gauge portal  for  SM extensions with $N_F$ generations of VLLs  of mass $M_F$ and in the representation $(0,\mathbf{2},\mathbf{1})$. Left panel: VLLs generate sufficient  uplift in comparison with SM running (full vs dashed lines).
     Right panel: critical surface of parameters in the $(N_F,M_F)$ plane, indicating whether the Planck scale vacuum  is  metastable (yellow), stable at or all the way up to the Planck scale (light vs dark blue), or plagued by a subplanckian Landau pole in $\alpha_2$ (red). The left panel model parameters are also indicated (white dot). The weak gauge portal extends substantially into the high mass and high multiplicity region. } 
    \label{fig:pNf5}
\end{figure*}

To illustrate how the gauge portal mechanism \eqref{quartic} operates quantitatively, we numerically  integrate the full 2-loop RG running,\,\cite{Hiller:2022rla,Litim:2020jvl} starting with the weak gauge portal characterized by $N_F$ new vector-like leptons (VLLs) of mass $M_F$ in the representation $(0,d_2,\mathbf{1})$.
A concrete example  $(N_F=5,{M_F}=1~{\rm TeV})$ is shown in \fig{pNf5} (left panel). 
The new VLLs induce a small uplift of the weak gauge coupling  (solid, green), just enough to stabilise the Higgs quartic (solid violet) along the  trajectory up to the Planck scale. 
In \fig{pNf5} (right panel), we perform a parameter scan in the $(N_F,M_F)$ plane to identify the ``critical surface'', i.e.~the BSM parameter regions where  the potential at the Planck scale is stable (blue),  metastable (yellow),  or plagued by a subplanckian Landau pole (red). 
If $M_F$ is too large and  $N_F$  too small,  there is not enough RG time  
to lift the instability, and the effects are ``too little too late''.  
On the other hand, if $N_F$ is too large and $M_F$  too small, the effects are too strong and predictivity is lost due to a Landau pole prior to the Planck scale. 
The sweet spot of SM extensions with stable vacua is situated in the wedge between the regions of metastability and Landau poles, which covers a wide range of multiplicities  $N_F$ and masses $M_F$. 
We also find regions where the uplift \eqref{quartic} ensures stability all the way up to $M_{\rm Pl}$ (dark blue), and regions where  ``squeezing''  dominates (light blue). 
We notice that there is no upper limit on $N_F$ nor $M_F$.
Larger $M_F$ implies that gauge couplings are much smaller at the matching scale. However, as long as subplanckian Landau poles are avoided, the smallness can be countered by larger $N_F$, which  allows $\alpha_2$ to grow fast enough to stabilise the Higgs. 

Interestingly, 
very similar results are found for the strong gauge portal,\,\cite{Hiller:2022rla,Gopalakrishna:2018uxn} even though  $\alpha_3$ contributions are loop-suppressed over  $\alpha_2$ contributions  \eqref{eq:quartic-enhance}.
Considering $N_F$ VLFs of mass $M_F$ in the representation $(0,\mathbf{1},d_3<{\bf 10})$  we again find wedges of stability, much like in \fig{pNf5}.
Unlike in the weak gauge portal, however, we now observe upper bounds on $N_F$ and on $M_F$, 
for example   $M_F \lesssim 10^6\,\text{TeV}$ and $2 \leq N_F \leq 18$ for vector-like quarks ($d_3=\bf{3}$).\cite{Hiller:2022rla} The reason for this is that  for too large $M_F$,   $\alpha_{3}$ and $\alpha_{t}$ at the matching scale are too small  to generate sufficient  uplift \eqref{quartic}. In fact, even if asymptotic freedom is lost (for large $N_F$), the growth of $\alpha_3$   and the  induced decrease of $\alpha_t$ are insufficient to generate  stability.

\begin{figure*}
    \centering
    \includegraphics[trim={2cm 0 2cm 0},clip,   width=.91\columnwidth]{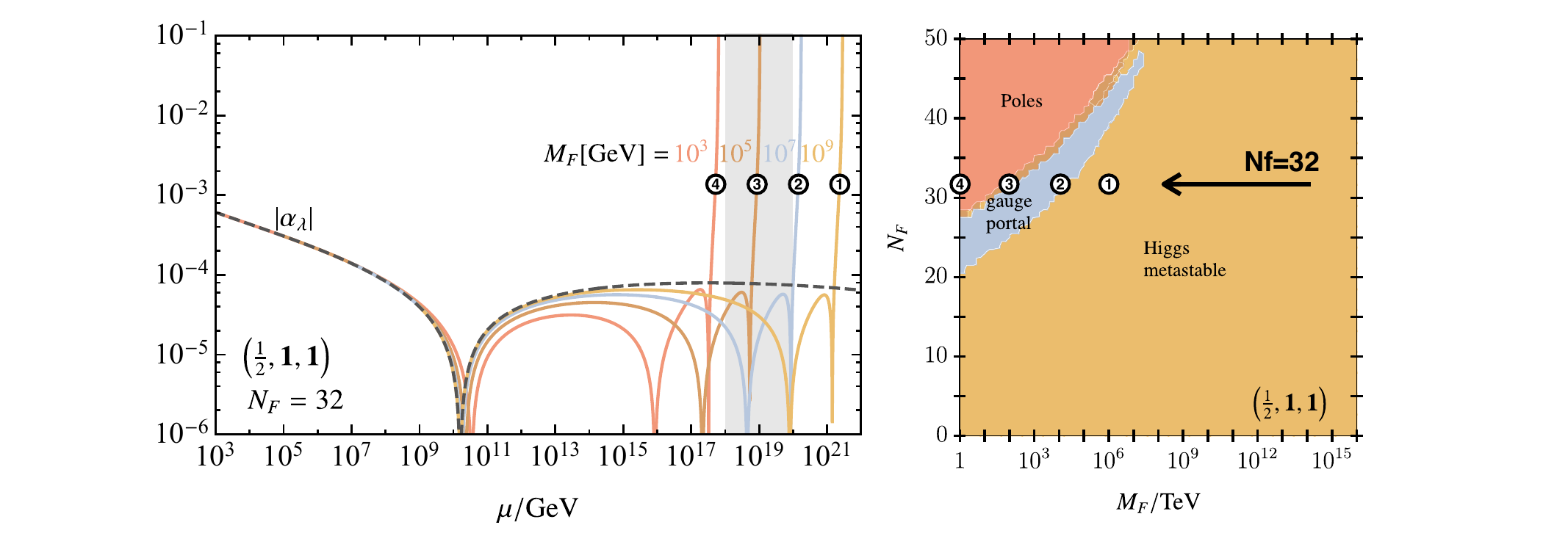}
\vskip-.5cm
    \caption{Illustration of the hypercharge  portal  for  SM extensions with $N_F$ generations of VLLs  of mass $M_F$ and in the representation $(\s012,\mathbf{1},\mathbf{1})$, showing the critical surface of parameters in the $(N_F,M_F)$ plane (right panel), colour-coding as in \fig{pNf5}. The left panel highlights the running of the BSM  Higgs quartic  in comparison to SM running (dashed line) for  $N_F=32$ and four different $M_F$, with corresponding
      dots also shown in the right panel.  We observe that the ``squeezing'' effect delivers  Planck-scale stability whereas the uplift is insufficient.}
    \label{fig:pNf32}
\end{figure*}

Finally, we consider the hypercharge  portal  characterized by $N_F$  VLLs of mass $M_F$ in the representation $(Y_F,\mathbf{1},\mathbf{1})$. 
Most interestingly, also the hypercharge portal is available, despite the looming Landau pole.
The  critical surface of parameters is shown  exemplarily  for models with $Y_F=\frac{1}{2}$ in \fig{pNf32} (right panel). Once more, we observe  a stability wedge between regions of   metastability and Landau poles.  Unlike the weak and strong portals, however, we do not find any region where $\alpha_\lambda>0$ all the way up to the Planck scale, showing that the uplift in  \eq{quartic-enhance} from hypercharge alone  is insufficient. 
Instead, stability  arises through ``squeezing'',
as illustrated in \fig{pNf32} (left panel) for 
$N_F=32$ and several values for $M_F$ (correspondingly highlighted by dots in the right panel). 
Given that $\alpha_1$ is  larger than $ \alpha_1^{\rm SM}$, we recognise the new $\alpha_\lambda$ trajectories in \fig{pNf32} as increasingly squeezed versions of the SM trajectory in \fig{pSM}.
Evidently, the effect is more pronounced for smaller $M_F$ as this triggers an earlier start of the accelerated $\alpha_1$ growth.  
If squeezing is too substantial, even the third sign change may arise prior to the Planck scale, typically around $\alpha_1 \gtrsim \text{few} \times 10^{-2}$, followed by an imminent Landau pole.

The hypercharge portal  disappears either by increasing $M_F$  thus leaving insufficient RG time for  squeezing to be operative, or by increasing $N_F$ leading to a subplanckian theory breakdown.
Increasing the hypercharge $|Y_F|$ causes the $N_F$-window to become narrower and to move towards lower $N_F$, and vice versa for $ Y_F \leftrightarrow N_F$.
Maximal hypercharges are achieved for smallest number of flavours, and
increasing $M_F$ for fixed $N_F$ enhances the overall range of viable $Y_F$.

\begin{figure*}[b]
    \centering
\vskip-.2cm
    \includegraphics[width=\columnwidth]{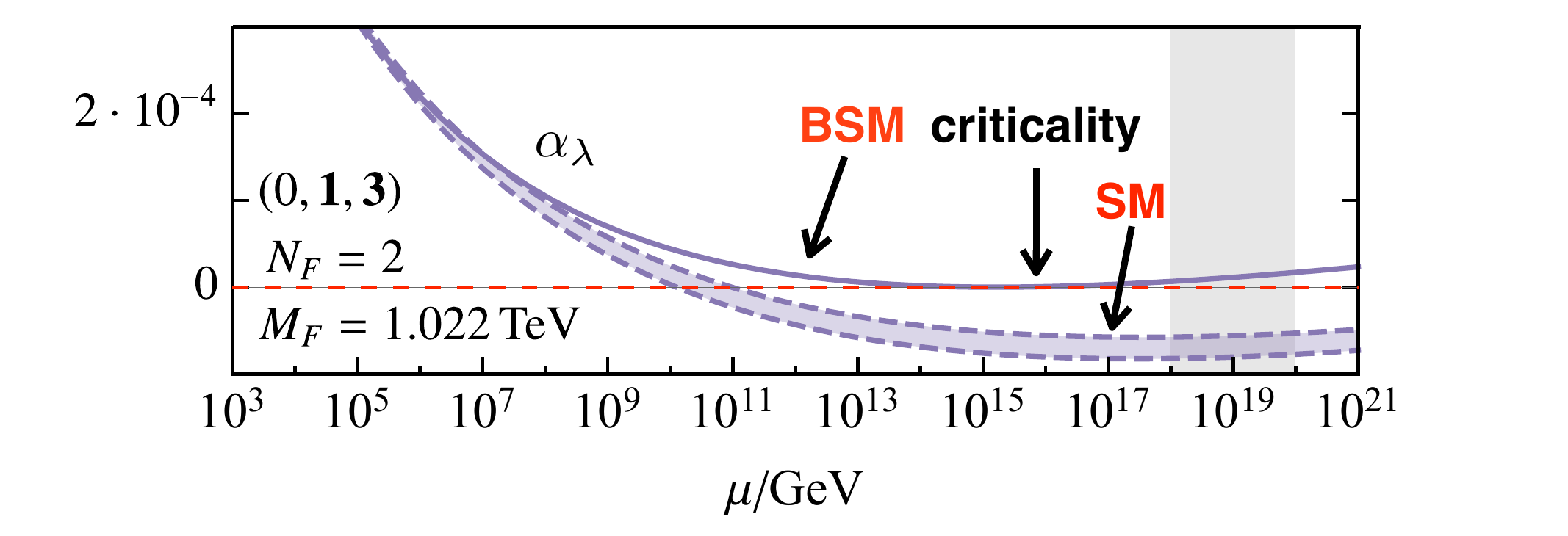}
\vskip-.2cm
    \caption{Comparison of BSM and SM Higgs criticality,  exemplarily for a SM extension with two vector-like quarks with charges $(0, \mathbf{1}, \mathbf{3})$ and mass $M_F=1.022$~TeV ($m_t = 172.76\,{\rm GeV}$). }    \label{fig:pcrit}
\end{figure*}

\section{How Critical is the Standard Model?}

It has been noticed previously~{}\cite{Buttazzo:2013uya} that the SM Higgs quartic is   near-critical, with 
$\beta_\lambda|_{\mu=M_{\rm Pl}}\approx 0$  and $\alpha_\lambda|_{\mu=M_{\rm Pl}}\approx -10^{-4}$,
reminiscent of a free RG fixed point at the Planck scale. It is    natural to ask whether SM extensions can be found where the quartic and its beta function vanish identically. 
We can answer this question to the affirmative:
the gauge portal mechanism allows us to find many suitable parameters $(N_F,M_F)$ and gauge charges $(Y,d_2,d_3)$ for VLFs such that 
$\alpha_\lambda$ achieves strict criticality, meaning  a double-zero at a scale $M_F\le \mu_{\rm crit}\le M_{\rm Pl}$, 
\begin{equation}\label{eq:crit}
\alpha_\lambda|_{\mu_{\rm crit}}= 0 \quad {\rm and}\quad \beta_\lambda|_{\mu_{\rm crit}}= 0\,.
\end{equation}
This is illustrated in \fig{pcrit}  where
the BSM Higgs quartic  $\alpha_\lambda$ remains positive  throughout and achieves   a double-zero  just above $\mu_{\rm crit}\approx 10^{15}$~GeV before  settling around $\alpha_\lambda\approx +10^{-5}$ at the Planck scale. As a result, many SM extensions can be found where the Higgs is as or more critical than in the SM, with the added benefit of stability rather than meta-stability.

\begin{figure*}
    \centering
 \includegraphics[width=.4\columnwidth]{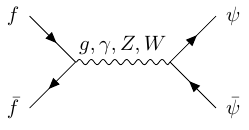}
\vskip-.2cm
    \caption{Pair-production  of vector-like fermions $\psi$ at pp and  $\ell\ell$ colliders, with $f$ indicating SM quarks or leptons.} 
    \label{fig:pProduction}
\end{figure*}

\section{Discussion}

 Lifting the instability of the SM vacuum has been put forward as a genuine ``bottom-up'' model building task. 
SM extensions by VLFs are particularly efficient for this because the gauge portal mechanism   enhances the Higgs quartic naturally, \eqref{quartic}.
Therefore, SM extensions with suitably charged VLLs and VLQs  over a large range of masses are   well-motivated. Moreover, searches at colliders and beyond, with broad signatures and production channels such as those indicated in \fig{pProduction},   are   strongly encouraged.
Further avenues towards stability arise in extensions with additional Yukawa or Higgs portals, giving   BSM parameter constraints analogous  to those shown in \fig{pNf5} and \fig{pNf32}.\,\cite{Hiller:2022rla,Hiller:2019mou,Bissmann:2020lge,Bause:2021prv}  
Settings with feeble or no Yukawas to the Higgs can be searched for in
$R$-hadron-signatures, or di-jets.\,\cite{Bond:2017wut}
 Models which allow for  flavourful Yukawas
give rise to flavourful constraints\,\cite{Hiller:2022rla}, and  allow to additionally address flavour anomalies.\,\cite{Hiller:2019mou,Bissmann:2020lge,Bause:2021prv}\\

\section*{Acknowledgments}
DFL  acknowledges support by the Science Technology and Facilities Council (STFC) through the  Consolidated Grant ST/T00102X/1.

\end{document}